\def\year{2022}\relax
\documentclass[letterpaper]{article} 
\usepackage{aaai22}  
\usepackage{times}  
\usepackage{helvet}  
\usepackage{courier}  
\usepackage[colorlinks=false, hidelinks]{hyperref}
\usepackage[square,numbers]{natbib}
\usepackage{graphicx} 
\usepackage{multirow}
\usepackage{amssymb}
\usepackage[left=1.75cm, right=1.75cm, top=2cm, bottom=2cm]{geometry}
\usepackage{natbib}  
\usepackage{pythonhighlight}
\usepackage{caption} 
\usepackage{xcolor} 
\DeclareCaptionStyle{ruled}{labelfont=normalfont,labelsep=colon,strut=off} 
\usepackage{xcolor}

\usepackage{algorithm}
\usepackage{caption}
\usepackage{subcaption}
\usepackage{listings}
\usepackage{multirow}
\usepackage{booktabs}
\usepackage{xcolor}
\usepackage{soul}
\usepackage{balance} 
\usepackage{algpseudocode}

\usepackage{newfloat}
\usepackage{listings}
\floatstyle{ruled}
\newfloat{listing}{tb}{lst}{}
\floatname{listing}{Listing}
\title{GOV-REK: Governed Reward Engineering Kernels for Designing Robust Multi-Agent Reinforcement Learning Systems}

\author{
    Ashish Rana \textsuperscript{\rm 1}, 
    Michael Oesterle \textsuperscript{\rm 1}, and
    Jannik Brinkmann \textsuperscript{\rm 1}
}
\affiliations{
    \textsuperscript{\rm 1} Institute for Enterprise Systems, University of Mannheim, Germany \\
    ashish.rana@students.uni-mannheim.de \\
    michael.oesterle@uni-mannheim.de \\
    jannik.brinkmann@uni-mannheim.de 
}

\usepackage{bibentry}

\begin{document}

\maketitle
\begin{abstract}

For multi-agent reinforcement learning systems (MARLS), the problem formulation generally involves investing massive reward engineering effort specific to a given problem.
However, this effort often cannot be translated to other problems; worse, it gets wasted when system dynamics change drastically.
This problem is further exacerbated in sparse reward scenarios, where a meaningful heuristic can assist in the policy convergence task.
We propose \textbf{GOV}erned \textbf{R}eward \textbf{E}ngineering \textbf{K}ernels (GOV-REK), which dynamically assign reward distributions to agents in MARLS during its learning stage.
We also introduce governance kernels, which exploit the underlying structure in either state or joint action space for assigning meaningful agent reward distributions.
During the agent learning stage, it iteratively explores different reward distribution configurations with a Hyperband-like algorithm to learn ideal agent reward models in a problem-agnostic manner.
Our experiments demonstrate that our meaningful reward priors robustly jumpstart the learning process for effectively learning different MARL problems.

\end{abstract}
\vspace{- 1 ex}

\section{Introduction}


Interactions formulated in MARLS are more intricate to learn for complicated scenarios at increasing scales and large numbers of agents \cite{tuyls2012multiagent, du2021survey}.
This problem is significantly exacerbated in multi-agent sparse scenarios as agent solution trajectories explode exponentially at large scales, and the reward signals are not dense enough to assist the learning process \cite{dann2015sample, jiang2018open}.
Many approaches have previously explored designing reward systems for single-agent and multi-agent settings \cite{ng1999policy, devlin2011theoretical}, either by exploiting domain knowledge or utilizing imitation learning \cite{brys2015policy, elbarbari2021ltlf} and ethic-based shaping learning novelties \cite{wu2018low}.
However, the reward engineering effort is problem-specific and often does not generalize to other MARL problems \cite{brys2015reinforcement, brys2014multi, lee2022marl}.
Hence, defining effective and robust reward signals for agents in MARL tasks in an automated and problem-agnostic manner is a challenging problem.

\begin{figure}
\centering
\includegraphics[width=0.325\textwidth]{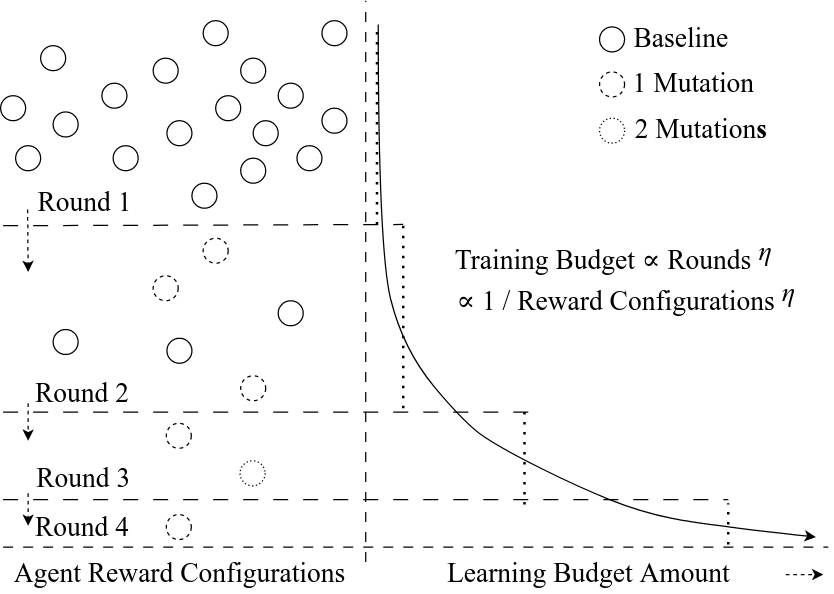}
\caption{The systematic reward configuration exploration with exponentially increasing training timestep budgets for learning.}
\label{fig:approach-intuition}
\end{figure}

\begin{figure*}
\centering
\includegraphics[width=0.925\textwidth]{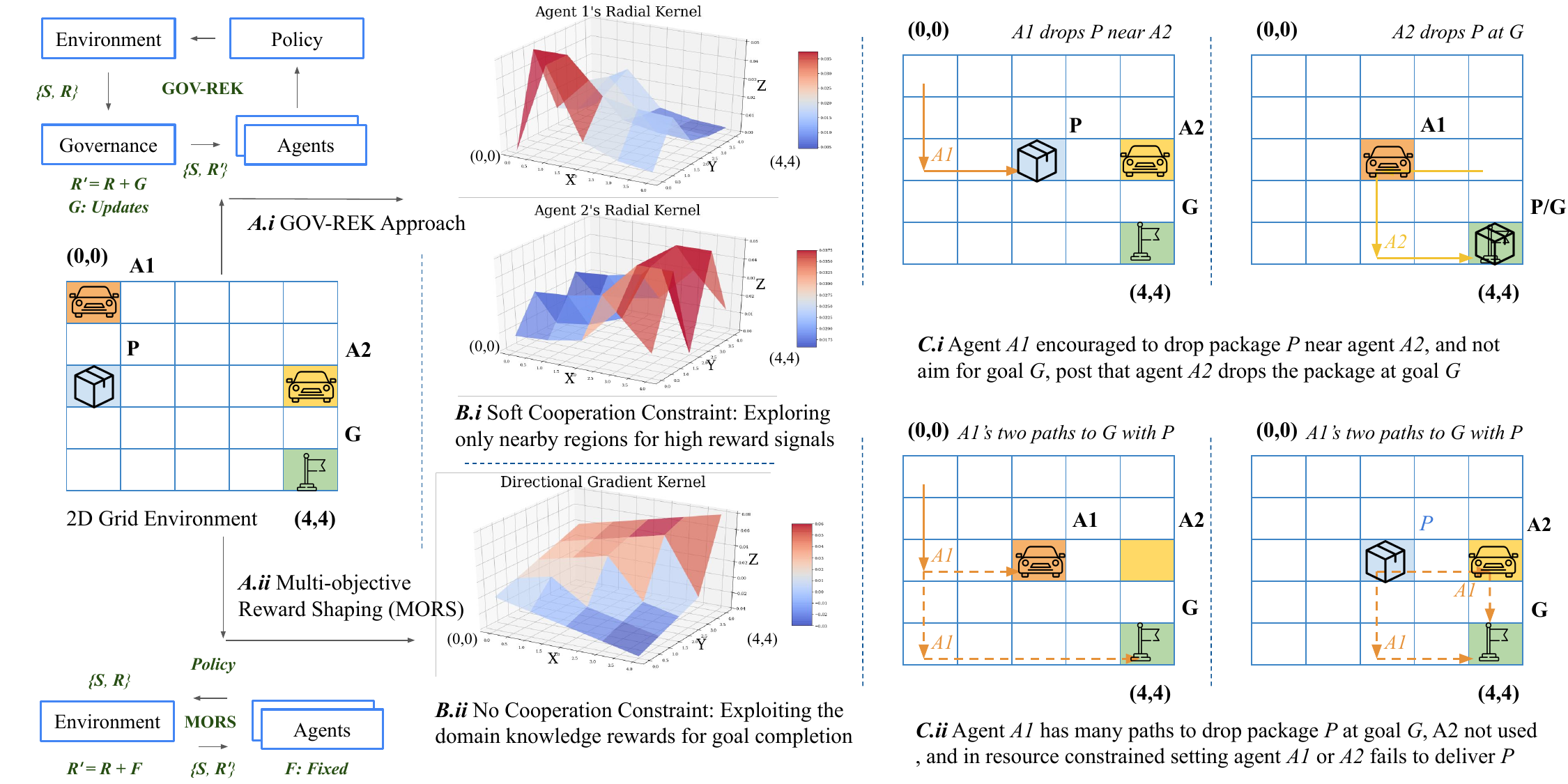}
\caption{The functional description of the governance layer for the sparse package delivery cooperative MARL problem.}
\label{fig:gov-func-desc}
\end{figure*}

Previously, architectural novelties introduced in Agent Temporal Attention (ATA) \cite{xiao2022agent}, Random Network Distillation (RND) \cite{burda2018exploration}, Never Give Up (NGU) \cite{badia2020never}, and Shared Experience Actor-Critic (SEAC) \cite{christianos2020shared} methods have successfully improved the learning behavior in reinforcement learning (RL) systems against sparse problems.
However, these approaches improve sample efficiency by introducing novelties like attention, curiosity, and experience sharing as part of the learning process rather than directly influencing agent motivations.
Our approach proposes an intermediary governance layer between agents and the environment, which directly incentivizes agents with additional rewards selected in an automated manner to improve the baseline RL algorithms.
In our proposed approach, we define ‘governance kernels’ for each agent, which are the reward distribution signals that generate similar additional rewards for similar states or joint actions depending on the MARL problem.
Also, for a plethora of machine learning (ML) and deep learning (DL) use cases, the hyperparameter optimization (HPO) algorithms \cite{bischl2023hyperparameter} like Successive Halving (SH) and Hyperband \cite{hutter2019automated} in a problem-agnostic manner have consistently found exemplary hyperparameter configurations \cite{young2015optimizing, bergstra2012random}.
Similarly, our proposed GOV-REK framework finds suitable reward models for agents by searching over different governance kernel configurations iteratively, where a repeated Hyperband-like algorithm generates the search plan to learn ideal governance kernels.
Further, the governance kernels are used in the GOV-REK framework as fundamental modules that can be superimposed and mutated across different model training rounds to incentivize agent cooperation.
Figure \ref{fig:approach-intuition} highlights our algorithm qualitatively, which executes for four rounds of SH and successively prunes out relatively worst reward configurations by factor $\eta$=3.
Further, as the round increases, the budget for each learning model increases exponentially by $\eta$=3 factor during successive rounds.
This increasing budget produces models with higher fidelities with each successive round with better configurations, and we also mutate a fraction of our selected best configurations.

Further Figure \ref{fig:gov-func-desc} demonstrates the functional working of the governance layer for the sparse package delivery problem, where the agent-specific reward is altered to \textit{R\textsubscript{i}} to \textit{R´\textsubscript{i}} by radial governance kernels.
Our proposed governance does not change state \textit{S\textsubscript{i}} and action \textit{A\textsubscript{i}} options for the agents, but the policy that chooses state \textit{S\textsubscript{i}} and action \textit{A\textsubscript{i}} might change with the newly added rewards.
For benchmarking we also highlight a comparative approach, Multi-Objective Reward Shaping (MORS), which assigns manually designed dense rewards for every subtask completion, like package exchange between agents, and proceeding closer toward the goal with dense directional reward gradients by exploiting domain knowledge \cite{brys2014multi}.
This governance layer is aware of all the rewards the agents receive under a complete or partial observability setting.
The governance coordinator exclusively influences each agent through additional agent-specific reward signals, but individual agent entities cannot directly change the governance-defined rewards.
For the system to achieve its shared objective, the governance evolves its reward distribution for agents based on changing agent behavior during learning, where the GOV-REK search plan executes different models with different governance kernel configurations.
The governance kernel selection criteria are flexible; it can either be a single objective, like maximizing rewards, or a multi-objective, like maximizing rewards and minimizing episode lengths.

Our first contribution is the inception of a dynamic inductive bias to explore topologically similar state or joint-action spaces for incentivizing cooperation amongst agents in MARLS.
Second, our proposed GOV-REK framework’s algorithm learns ideal agent reward models by conducting an iterative search over our proposed problem-agostic reward distributions.
Further, we also demonstrate that our proposed method successfully applies to different MARL problem task configurations in an automated manner.

\section{Related Work}
\label{related-work}

\begin{figure*}
\centering
\includegraphics[width=.975\textwidth]{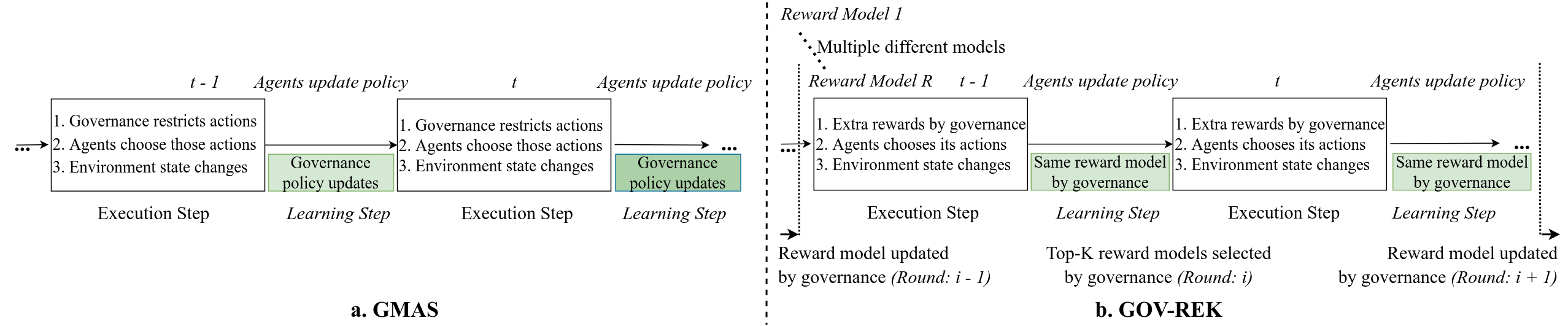}
\caption{The sequence of execution and learning steps in GMAS and GOV-REK approaches.}
\label{fig:governance-functionality}
\end{figure*}

Defining a good MARL goal is a challenging objective, where expected agent rewards need to be jointly maximized in a completely observable or partially observable setting.
The defined rewards for a given goal must stabilize the agent's learning behavior while adapting to changing dynamics in the environment.
The stability convergence requirement ensures the stationary policy convergence, and the adaptability constraint ensures no performance detriment with evolving policies of other agents, provided agents are rational \cite{bowling2001rational, chalkiadakis2003multiagent}. 
Further, for training MARLS, the Centralized Training Centralized Execution (CTCE) paradigm optimizes the joint policy for agents together, and the Centralized Training Decentralized Execution (CTDE) paradigm agents maintain separate policies but exchange information during training \cite{zhao2020sim, gronauer2022multi, du2021survey}.
In this study, we train our MARLS with CTCE in a fully observable setting and CTDE in a partially observable setting against two different MARL problems to quantify the scalability and adaptability performance aspects.
Previously, architectures utilizing additional novelties like imitation-based learning \cite{schaal1996learning, brys2015reinforcement}, curiosity-driven learning \cite{bellemare2016unifying, ostrovski2017count}, curriculum learning \cite{baker2019emergent}, self- or temporal-attention \cite{iqbal2019actor, jiang2018learning}, and evolutionary learning \cite{long2020evolutionary} have shown efficacy for a wide range of RL problems.
With our proposed GOV-REK framework, we focus on improving the performance of existing baseline algorithms with an additional coordinating governance layer that primarily alters agent incentives to achieve convergence.

For defining meaningful agent motivation, reward-shaping has been widely studied in the past, where this approach has proven its efficacy for achieving faster convergence \cite{eschmann2021reward, wirth2017survey}.
Also, incorporating other novel mechanisms, like learning ethical human behavior demonstrations \cite{wu2018low}, multi-objective reward shaping (MORS) \cite{brys2014multi}, additional rewards for sub-goal completion \cite{harutyunyan2015expressing}, and context-sensitive rewards for agents \cite{brys2015reinforcement}, have shown further improvements.
However, reward-shaping agents are often susceptible to falling under continuous positive reward cycle traps, especially for sparse environments where additional rewards can dominate the accurate underlying reward model.
Formally, the reward function for the underlying Markov Decision Process (MDP) can be modified with the relation \textit{$R'$ = $R + F$}, where \textit{F(s, s, $s'$)} is the additional transition reward model, and \textit{$f_t$} is defined analogously to \textit{$r_t$} in a temporal setting.
The Potential Based Reward Shaping (PBRS) maintains a potential function \textit{$\Phi:\mathcal{S}$ $\rightarrow \mathbb{R}$} is a necessary and sufficient constraint designed for policy invariance which applies to MARLS as well \cite{devlin2012dynamic, devlin2011theoretical}.
This relation can further be modified to incorporate the temporal element given by $F(s, t, s', t') = \gamma \Phi(s', t') - \Phi(s,t)$, where $\gamma$ denotes the discount factor.
Furthermore, our GOV-REK approach restricts all our agent solution trajectories to always satisfy PBRS constraints by using only normalized reward distributions as additional reward signals.

The task of finding optimal strategies or policies in MARL systems is still an open challenge \cite{zhang2021multi}.
To mitigate this problem, researchers have proposed a paradigm where agents are provided with assistive information for learning.
Approaches, like Environment-Mediated Multi-Agent Systems (EMMAS) \cite{weyns2008engineering}, Electronic Institutions (EI) \cite{esteva2001formal}, and Normative Multi-Agent Systems (NorMAS) \cite{conte1999agents, DBLP:conf/icaart/Neufeld22} generally employ a restrictive strategy to limit original solution policy space for empirically achieving faster convergence.
Further, the Autonomic Electronic Institutions (AEI) approach dynamically evolves these constraints to achieve even faster convergence \cite{bou2006towards}.
As shown in Figure \ref{fig:governance-functionality}, the Governed Multi-Agent System (GMAS) approach queries every execution step to obtain permissible actions, and the learning happens between those steps \cite{oesterle2022self}.
Also, in each learning step, the governance optimizes its learning policy for maximizing the system objective while evolving the action-space constraints.
The black-box ANN agents also update their policies at each learning step, where the agents are not part of the governance.
All the above-discussed approaches strictly constrain the agent action spaces, which might be suboptimal when a massive exploration of the joint state-action space is needed, like in sparse reward problems.
However, with our proposed approach, the additional reward signals introduce only soft constraints on agent exploration behavior, which prohibits strictly restricting the exploration capacity of agents.
Second, our reward models are evolved in a more stable manner, where only better reward models are selected after each round of significant model training as highlighted in Figure \ref{fig:governance-functionality}.

\begin{figure*}
\centering
\includegraphics[width=0.89\textwidth]{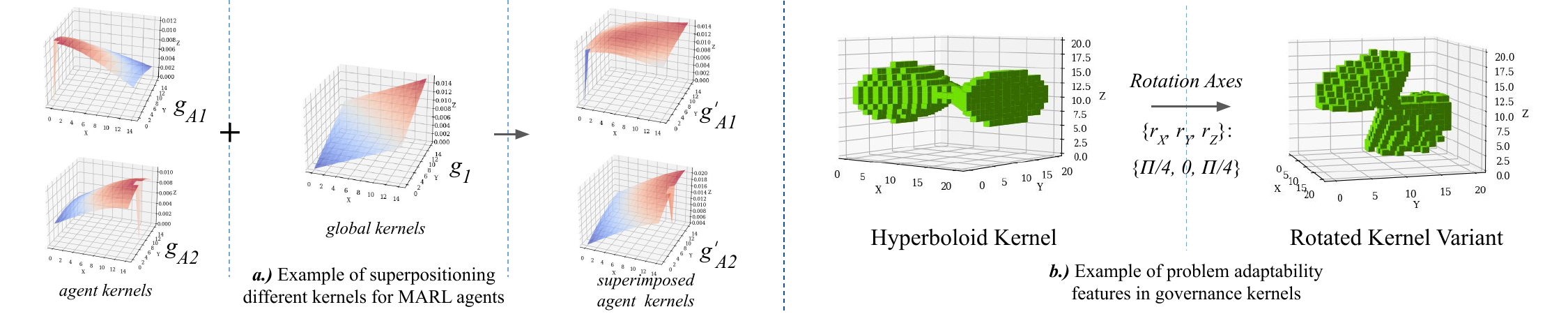}
\caption{The different ways in which different governance kernels are superimposable and adaptable as agent-specific and agent-agnostic reward distributions.}
\label{fig:gov-kernel-adaptability}
\end{figure*}

The non-parametric approaches like Gaussian Process Regression (GPR) use Gaussian kernel priors to perform the regression task.
Also, the kernel's choice determines the GP model's generalization properties in the GPR modeling task \cite{duvenaud2014automatic}.
Here, we also focus on defining meaningful prior reward distribution for agents for incentivizing agent cooperation behavior in sparse scenarios.
Also, HPO methods have proved highly effective for finding suitable hyperparameter configurations of prediction and modeling problems \cite{bischl2023hyperparameter}.
For example, SH explores the hyperparameter space from low fidelity configurations with a lower training budget first, and with each increasing round, doubles the training budget by removing underperforming configurations \cite{hutter2019automated}.
With increasing rounds, half of the active configurations with more enormous losses are eliminated, and their budget is equally assigned to the existing configurations.
Further, the Hyperband algorithm carries out multiple SH rounds across parallel brackets to train a variety of hyperparameter configurations with different training budgets to explore more configurations systematically in a problem-agnostic manner \cite{hutter2019automated}.
Based on our learnings from Hyperband’s iterative exploration methodology, we also iterate over different governance kernels in a problem-agnostic manner in our proposed approach using a repeated Hyperband-based search.


\section{Approach Formulation}
\label{approach-formulation}

Computationally, optimizing the joint MARL policy while parallelly searching for ideal reward models for agents is practically infeasible for complex scenarios. 
We utilize the similarities in state and joint-action spaces to define simplistic reward model distributions as a function of spatial topology in the MARL problem tasks.
Hence, simplistic reward models denoted as governance kernels are defined only based on the geometric similarities in the state or joint action spaces but not on state-action transitions.

\subsection{Governance Kernels}

The generalization of MDP formulation for RL tasks is the stochastic game (SG) formulation in MARL settings, where it is formally defined by the tuple \textit{$<S, A_1,$ ..., $A_n, f, R_1,$ …, $R_n >$} \cite{bacsar1998dynamic}. 
For \textit{n} agents, \textit{S} represents a discrete set of states, \textit{$A_i \in$ \{1, …, n \}} represents the discrete action set.
The discrete actions are chosen from the discrete values range of \textit{k} actions, which further generates the joint action space \textit{$\mathcal{A} = A_1 \times$ … $\times A_n$}.
The function \textit{f: S $\times \mathcal{A} \times$  S} represents the state transition probability, and reward functions for agents \textit{$R_i \in$ \{1, …, n \}} are expressed as \textit{$R_i:$ S $\times \mathcal{A} \times$ S $\rightarrow \mathbb{R}$}.
Further, the individual agent policies \textit{$\pi_i$: S $\times \mathcal{A} \rightarrow$ [0,1]} together form the joint policy \textit{$\pi$}, and POSG is a more generalized version of SG where the underlying states are unknown.
MARL algorithms can be implemented on static games \textit{(stateless)} like the social-dilemma problem where the state set \textit{S = $\phi$} and reward only depends on the joint actions \textit{$R_i: \mathcal{A}$ $\rightarrow \mathbb{R}$} \cite{leibo2017multi}.
Also, MARL algorithms apply to the stage games where S $\neq$ $\phi$. Our experiments demonstrate the utility of defining simplistic reward models that exploit geometrical spatial properties in state space in stage games and joint-action space in static games.

\begin{table*}[t]
\centering
\caption{The mathematical formulation of common Gaussian kernels and 3D-surface functions as sample governance kernels for 2D and 3D spatial MARL tasks respectively.}
     \label{tab:governance-kernel-functions}
    \begin{center}
    \scalebox{0.9}{
        \begin{tabular}{llllll}
        \toprule
            \multicolumn{3}{c}{Governance Kernels for 2D spaces} & \multicolumn{3}{c}{Governance Kernels for 3D spaces} \\
            \midrule 
            Name & Equation & Kernel & Name & Equation & Kernel \\
        \midrule 
             Linear Kernel & \textit{$\sigma^2(x-c)(x'-c)$} & \begin{minipage}{.1\textwidth}
             \includegraphics[width=11mm, height=11mm]{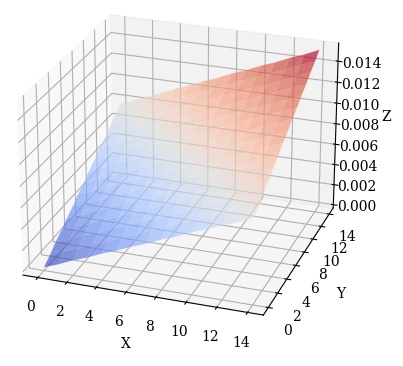}
             \end{minipage} &
             Diagonal Kernel & \textit{$x = y = z$} & \begin{minipage}{.1\textwidth}
             \includegraphics[width=11mm, height=11mm]{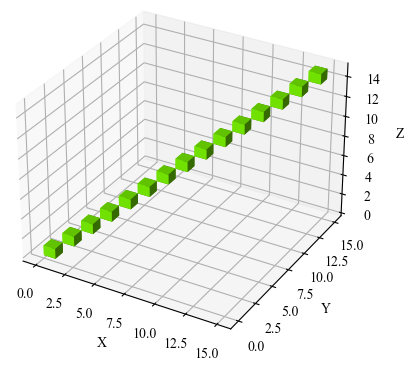}
             \end{minipage}\\
             Periodic Kernel & \textit{$\sigma^2 \exp\Bigl(\frac{-2sin^2(\pi|x-x'|^2/p)}{l^2}\Bigr)$} & \begin{minipage}{.1\textwidth}
             \includegraphics[width=11mm, height=11mm]{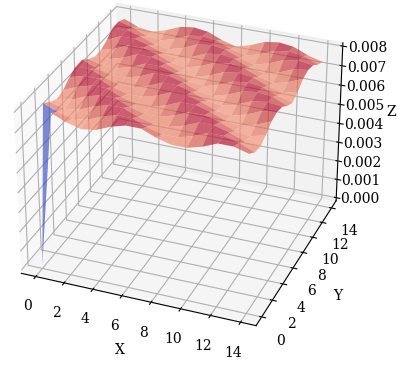}
             \end{minipage} &
             Ellipsoid Kernel & \textit{$\frac{x^2}{a^2} + \frac{y^2}{b^2} + \frac{z^2}{c^2} = 1$} & \begin{minipage}{.1\textwidth}
             \includegraphics[width=11mm, height=11mm]{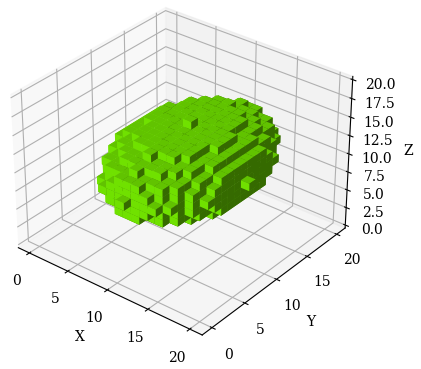}
             \end{minipage}\\
             Squared Exponential & \textit{$\sigma^2 \exp\Bigl(\frac{-(x-x')^2}{2l^2}\Bigr)$} & \begin{minipage}{.1\textwidth}
             \includegraphics[width=11mm, height=11mm]{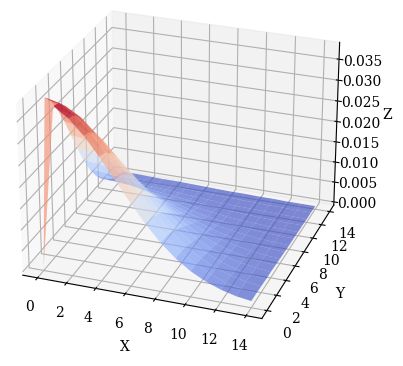}
             \end{minipage} &
             Hyperboloid Kernel & \textit{$\frac{x^2}{a^2} + \frac{y^2}{b^2} - \frac{z^2}{c^2} = 1$} & \begin{minipage}{.1\textwidth}
             \includegraphics[width=11mm, height=11mm]{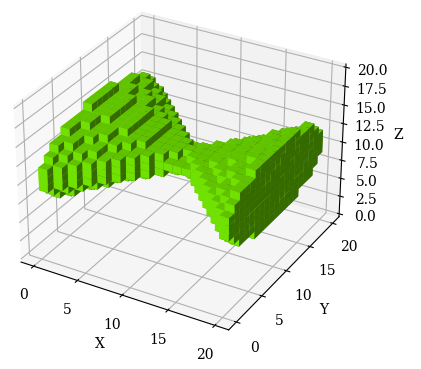}
             \end{minipage} \\
        \bottomrule
        \\
       \end{tabular}}
     \end{center}
\end{table*}

\begin{equation} \label{eq:def-governance-kernel}
\textit{$g_i = \sigma^2 \kappa(s_{a_i}, s'_{a_i}) + \xi; G = \sigma^2 \kappa(s, s') + \xi; $}
\end{equation}

For defining governance kernels, which are the simplified reward models proposed as part of the GOV-REK framework, we assume that \textit{$E_a$ $\bigl[$R(s, a, $s')\bigl]$ $\rightarrow$ $R'$(s, $s'$)} holds.
This exploration expectation assumption means that the underlying learning algorithm is highly curious to select diverse actions, where all the relevant solution trajectories between the state-action transition pairs are explored.
Hence, mathematically, we take an expectation with respect to the explored actions from the solution trajectories to define our reward models only as a function of state similarity, and we extend our results to joint-action spaces as well.
The performance of our solution directly depends on the enhancement provided by the governance kernel rewards $G_R$ for sampling all relevant solution trajectories with additional assistance from the learning algorithm.
Nevertheless, it allows us to define governance kernels independent of agent transitions, and this is expressed by the relations $r'_i$ = $r_i +$ $g_{r, i}$ and $R'$ = $R +$ $G_{r}$ for agent-specific and agent-agnostic kernels respectively.
The agent-specific kernels are defined considering the initial agent's spatial locations, whereas the agent-agnostic kernels are independent of the agent's initial location.
Therefore, the convergence for MARL tasks is accelerated with suitable governance kernel priors and high exploration curiosity.
In its generalized mathematical form, we express our governance kernels with the equation \ref{eq:def-governance-kernel} as agent-specific and agent-agnostic non-parametric variations, respectively.
Here, the kernel function is represented by $\kappa$, $\sigma$ upscales or downscales function values, $(s_{a_i},$ $s'_{a_i})$ or (s, $s'$) quantifies the magnitude variation between agent-specific or agent-agnostic states, and $\xi$ represents the noise in the kernel function.

\begin{figure*}
\centering
\includegraphics[width=0.87\textwidth]{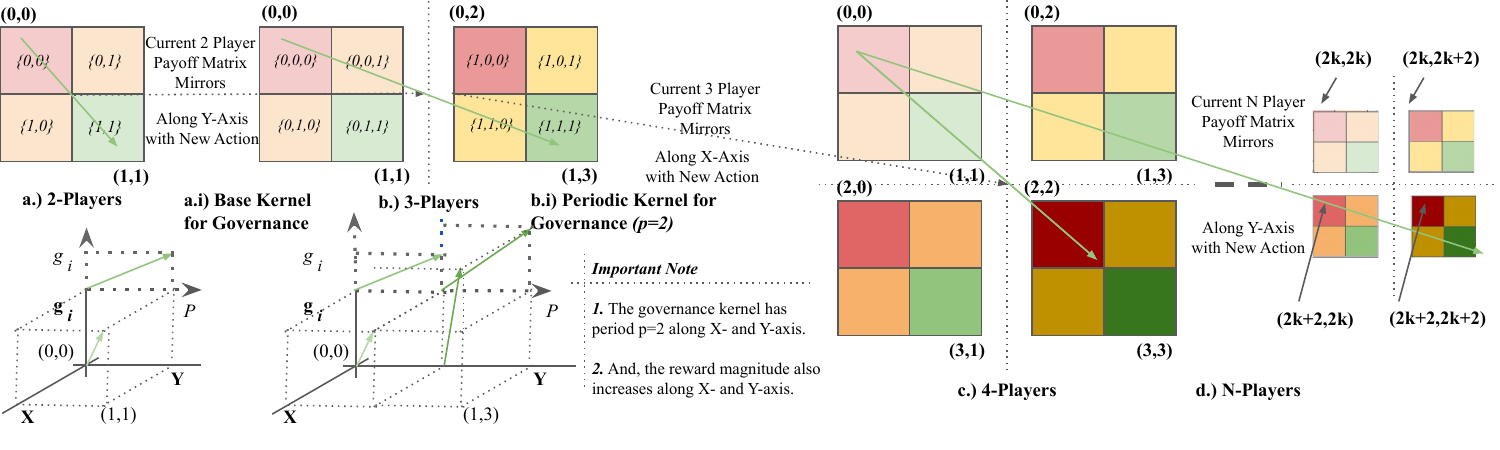}
\caption{The increasing periodic governance kernel configuration for the non-spatial social dilemma problems.}
\label{fig:non-spatial-intuition}
\end{figure*}

Conceptually, kernels are widely used in GPR with Gaussian kernels to generate model function estimates, and the non-parametric Gaussian kernels are usable in modeling complex functions for prediction tasks \cite{duvenaud2014automatic}.
Also, similar to Gaussian kernels, these governance kernels can be superimposed over one another to generate even more complex reward models.
Table \ref{tab:governance-kernel-functions} highlights some sample governance kernels that resemble popular Gaussian kernels, where the radial scale, periodicity, and magnitude are tunable for these kernels.
The governance kernels are building blocks of the GOV-REK framework, which generates problem-agnostic reward models.
It further provides orientation adaptability with superimposition capabilities to define spatially flexible and complex reward models as highlighted in Figure \ref{fig:gov-kernel-adaptability}.
To extend the package delivery problem into a 3D environment with drones, we use another set of governance kernels with volumetric surfaces and conical sections, providing easy interpretability \cite{gomes2009implicit, benny1922plane}.
Mathematically, the formal definitions of some sample surface governance kernels are provided in Table \ref{tab:governance-kernel-functions}.
The usage of these geometrical surfaces further highlights the simplicity of these kernels and their extensibility to non-Guassian governance kernels.
Further, we also assign similar governance kernels to joint-action spaces in the reward payoff matrix for an extension to static games, which are defined without explicit states.
For non-spatial problems, like the social dilemma problems, we demonstrate that their joint action reward payoff matrix also consists of spatial trends, as shown in Figure \ref{fig:non-spatial-intuition}.
Specifically from Figure \ref{fig:non-spatial-intuition}, we observe that a periodic governance kernel with a specific periodicity superimposed with a directional gradient governance kernel can further encourage agents to cooperate.
Generally, stage SGs and static SGs enclose a geometric state space or joint-action space representation property.
Furthermore, the proposed governance kernels are applied over this geometric topology to bias the agent behavior to assist the learning algorithm.

\subsection{The GOV-REK Framework}

The GOV-REK framework uses repeated Hyperband executions and iteratively manipulates governance kernel configurations to find suitable agent reward models.
For imposing PBRS consistency constraints, we normalize all the reward values associated with each state or joint-action element.
It is done by dividing each scaler reward element by the sum of total additional rewards added by that governance kernel multiplied by the number of agents.
The pseudocode algorithm \ref{alg:gov-rek-algorithm}, highlights our proposed methodology of repeated execution of Hyperband-like $N_r$ rounds, where $T$ represents the total SH training bracket budget.
The $total\_budgets$ stores decreasing SH bracket budgets for repeated Hyperband round execution, $\eta$ represents the multiplication factor which increases the training budget and decreases the number of configurations after each SH round.
In algorithm \ref{alg:gov-rek-algorithm}, the first training loop defines the number of Hyperband-like round repetitions, and in the later Hyperband-like rounds, the selected best kernels from each SH bracket are mutated with the factor $m=.5$.
Further, the second training loop defines the bracket training budget $R$, maximum brackets in current round $s_{max}$, and total Hyperband plan budget $B$ encompassing multiple SH brackets.

\begin{algorithm}
\caption{Algorithm pseudocode for the GOV-REK framework} \label{alg:gov-rek-algorithm}
\begin{algorithmic}
  \State $Input: T, N_r, \eta, t_k$
  \State $Set: total\_budgets = reverse\_sort([T \times \frac{i}{N_r}$ $in$ $(N_r)])$
  \State $Set: global\_top\_conf\_set_{gov} = []$
  \For{\textit{$R$ $in$ $total\_budgets$}}
    \State \textit{$Set: s_{max} = \lfloor log_{\eta}(R) \rfloor, B = (s_{max}+1)R $}
    \State \textit{$Set: round\_top\_conf\_set_{gov} = []$}
    \For{\textit{$s \in \{s_{max}, s_{max}-1, ..., 1, 0\} $}}
      \State \textit{$Set :$ $n_{gov} = \lceil \frac{B}{R} \frac{\eta^s}{s+1} \rceil; r_{gov} = R\eta^{-s};$}
      \State \textit{$Set :$ bracket\_conf\_set$_{gov}$ = []}
      \State \textit{get\_conf\_set$_{gov}$ = get\_governance\_kernels($n_{gov}$)}
      \For{\textit{ $j \in \{ 0, ..., s\}$ }}
        \State \textit{$Set : n_{gov,j} = \lfloor n_{gov} \times \eta^{-j} \rfloor; r_{gov,j} = r_{gov} \times \eta^{i} $ } 
        \State \textit{$L =\{get\_trained\_model\_metrics($}
        \State \textit{\hskip4.0em: $conf_{gov}, r_{gov,j})$}
        \State \textit{ \hskip2.0em: $\in$ conf\_set$_{gov}\}$}
        \State \textit{$bracket\_conf\_set_{gov}.add($}
        \State \textit{ \hskip2.0em $top\_configs(t_k, L, m=.5, s=.5))$}
      \EndFor
      \State \textit{round\_top\_conf\_set$_{gov}$.add(}
      \State \textit{ \hskip2.0em top\_configs($t_k$, bracket\_conf\_set$_{gov}$, L))}
    \EndFor
    \State \textit{global\_top\_conf\_set$_{gov}$.add(}
    \State \textit{  \hskip2.0em top\_configs($t_k$, round\_conf\_set$_{gov}$, L))}
  \EndFor
  \State \textit{Return: global\_top\_conf\_set$_{gov}$ trained models}
\end{algorithmic}
\end{algorithm}

Finally, the last training loop defines the different governance kernel configurations and their corresponding budgets for the SH bracket.
Furthermore, after the parallel execution of multiple SH rounds across different brackets, the top governance kernels are selected.
These best-performing governance kernels with the best accumulated average reward and minimum average episode length values are passed to later rounds.
Further, they are genetically mutated and superimposed with other governance kernels with mutation probability factor $m=.5$ and superimposition probability factor $s=.5$, respectively.
The total accumulated reward for the governed MARL includes additional governance kernel rewards for all agents.
Also, the merged kernels are normalized again to avoid violating PBRS constraints after governance kernel superimposition.
If the mutated or superimposed governance kernels decrease the model performance objectives, the previous top-performing kernels for agents get selected by default.

\section{Experiments}
\label{experiments}

In our experimentation, we test the comparative performance, robustness, and scalability of the GOV-REK framework on a 2D-grid road and 3D-grid drone environment in a fully observable CTCE setting.
Further, to test the efficacy and adaptability of our approach, we extend our proposed GOV-REK framework onto the social dilemma problem in a partially observable CTDE setting.
Also, for the spatial package delivery task, the governance kernels are defined over the state space, and contrastively, for the non-spatial social dilemma problem, the governance kernels are defined over the joint-action space.
For the package delivery problem, effectively, both resource constraint agents must cooperate to deliver a package to the goal location to receive the only goal reward.
Whereas in the social dilemma problem, we define two scenarios, where partial cooperation amongst agents yields partial rewards in the first scenario, and in the second scenario, only complete cooperation amongst the agents yields the only cooperation reward.
The proposed governance layer alters the net reward for the system for the training and inference stages.
Also, the new average expected reward includes the additional reward provided by the agent governance kernels.

We select the baseline Proximal Policy Optimization (PPO) implementations with default hyperparameters from Stable Baselines3 \cite{stable-baselines3} and RLlib \cite{liang2018rllib} packages for CTCE and CTDE training respectively.
Further, each of the learning curves for our experiment task results from the average of five different executions having different random seeds, and 95 \% confidence interval ranges are also plotted to quantify the performance uncertainty.
Our execution of the GOV-REK plan demonstrates that the superimposition of agent-specific squared exponential and agent-agnostic linear governance kernels works best for the $5 \times 5$ 2D-grid road setting.
Similarly, for the grid drone environment, a combination of agent-agnostic hyperboloid and diagonal surface governance kernels works best for the $3 \times 3$ 3D-grid drone setting.
Finally, for the social dilemma problem, a combination of linear and periodic governance kernels progressively directs the agents toward higher cooperation rates.
In our experimentation tasks, we analyze the impact of random perturbations, increasing scale and complexity, and symmetry in cooperation contributions for the governance kernels selected by the GOV-REK framework.

\begin{figure*}
\centering
\includegraphics[width=.975\textwidth]{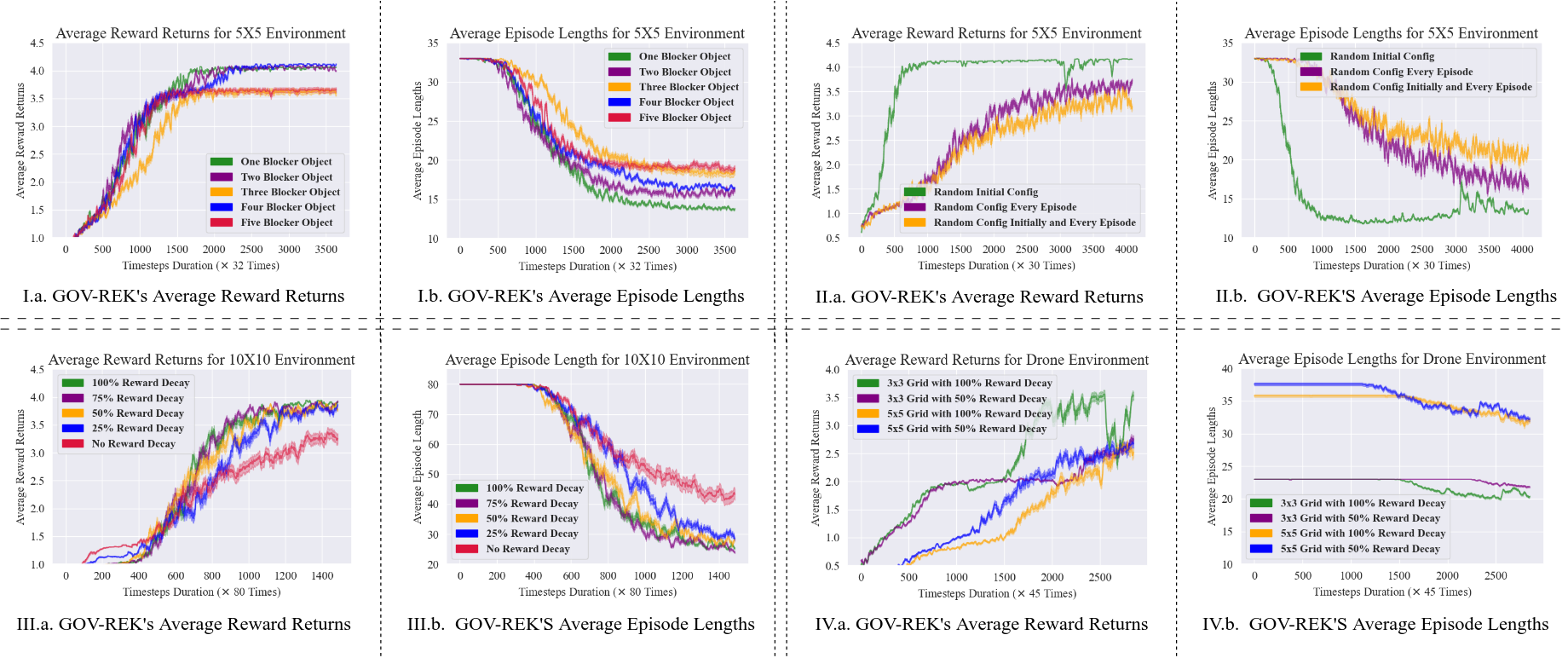}
\caption{The GOV-REK experiment summary measures average reward returns and episode lengths for the governed MARLS to quantify i.) robustness against increasing path blocker objects in $5 \times 5$ 2D-grid road environments, ii.) robustness against different randomization perturbations in $5 \times 5$ 2D-grid road environment configurations, iii.) scalability performance with different reward decays in $10 \times 10$ 2D-grid road environments., iv.) scalability performance with different reward decays in $3 \times 3$ and $5 \times 5$ 3D-grid drone environments.}
\label{fig:gov-rek-grid-env-experiments}
\end{figure*}

\subsection{Robustness Analysis}

The I and II subfigures in Figure \ref{fig:social-dilemma-result-summary} demonstrate the robustness of the GOV-REK framework against randomized perturbations in environment configurations and solution trajectory blocker objects.
We observe that the governed MARLS trained for 120K timesteps are generally robust to an increasing number of blocker objects. However, the average episode length increases as the number of blockers increases.
Further, relatively larger average episode lengths for the per-episode randomized environments highlight non-optimal solution trajectory selection behavior and slower package deliveries.
The increased average episode length behavior is attributed to agents selecting sub-optimal trajectories to tackle the randomized behavior added by changing configurations and blocker object locations.
The selected governance kernels by the GOV-REK framework are not optimized for changing initial and per-episode random configurations.
However, the selected governance kernels still demonstrate robustness against these random environment perturbations.

\begin{figure}
\centering
\includegraphics[width=.46\textwidth]{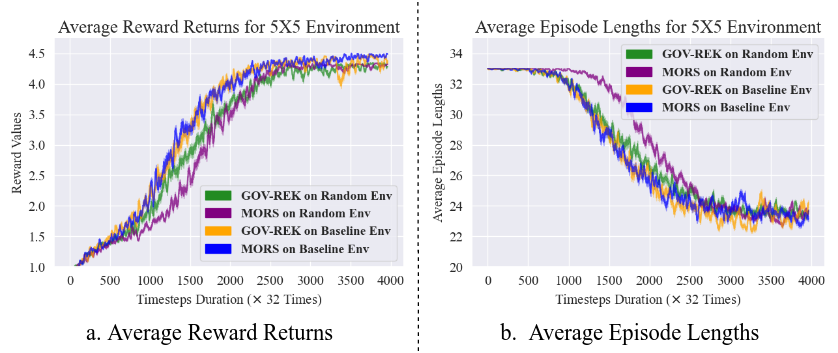}
\caption{The comparison between the GOV-REK and MORS approaches for a $5 \times 5$ size 2D-grid road environment in different configurations.}
\label{fig:gov-rek-vs-mors-obj}
\end{figure}

\subsection{Scalability Analysis}

\begin{figure}
\centering
\includegraphics[width=0.46\textwidth]{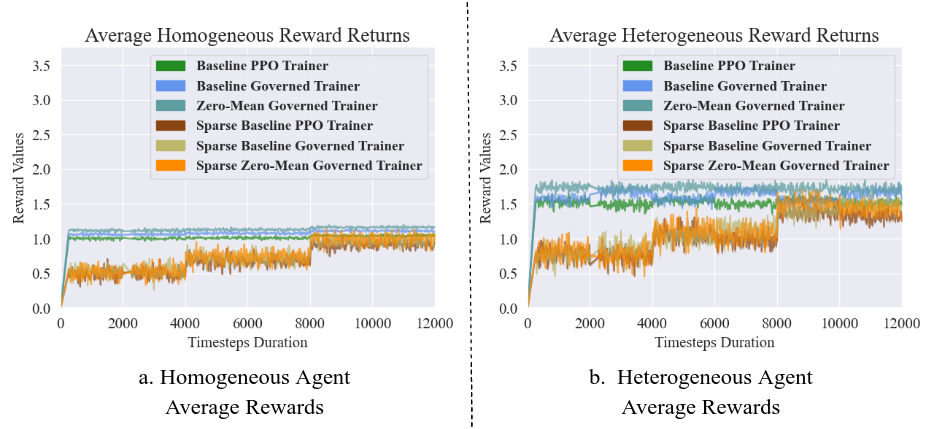}
\caption{The average expected reward returns for a single agent in homogeneous and heterogeneous agent systems for baseline and sparse social dilemma problems.}
\label{fig:social-dilemma-result-summary}
\end{figure}

Theoretically, for each agent in the 2D-grid road and 3D-grid drone environments, the number of possible solution trajectories are of the factorial order given by $(l+w-2)!$ $/$ $(l-1)!$$(w-1)!$ and $(l+w+h-3)!$ $/$ $(l-1)!$$(w-1)!$$(h-1)!$ respectively, where $w$, $l$, and $h$ stands for weight, length, and height respectively.
Therefore, with increasing linear scale, the problem complexity increases in factorial complexity, and adding the second agent further adds to the MARL task complexity.
Our approach, in its nascent form, is partially unable to assist the baseline PPO algorithm in exploring diverse solution trajectories optimally, which leads to exploration expectation assumption violation. 
To mitigate this solution trajectory sampling issue, we introduce decaying governance kernels that reduce the future reward value associated with the state to the given fractional amount when an agent visits that state.
The baseline governance kernels bias the agent's curiosity to explore local regions more thoroughly, but decaying governance kernels encourage agents to explore more global and diverse solution trajectories.

\begin{figure}
\centering
\includegraphics[width=0.45\textwidth]{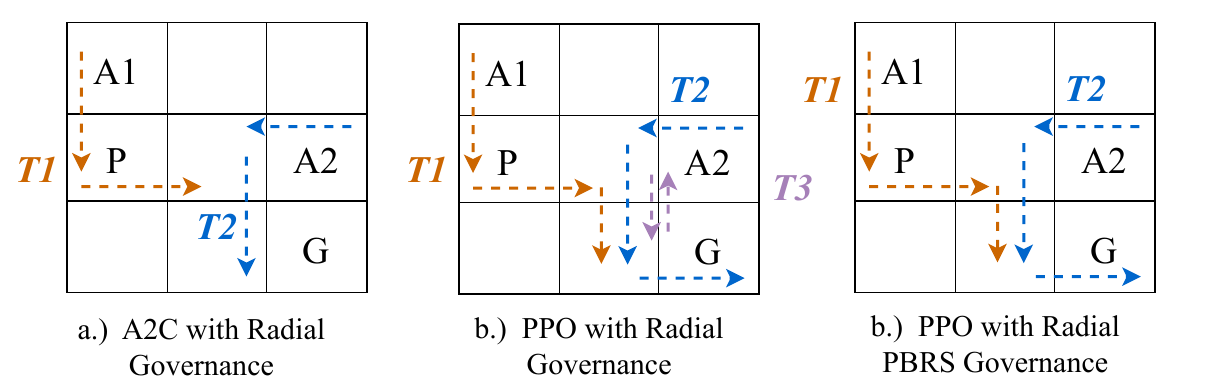}
\caption{Different trajectories followed by governed agents with A2C and PPO learning algorithms with governance.}
\label{fig:3x3-qualitative-trajectory-plot} 
\end{figure}

Further, III and IV subfigures in Figure \ref{fig:social-dilemma-result-summary} demonstrate the capabilities of the GOV-REK framework to operate on larger $10 \times 10$ 2D-grid road and $5 \times 5$ 3D-grid drone environments effectively.
For a 2D-grid road environment, we observe that governance kernels are more effective with higher decay rates, leading to faster convergence.
Second, for a 3D-grid drone environment, the efficiency again increases with higher decay rates, but the average episode length reduction is relatively less.
We attribute this relative performance depreciation to the simplistic nature of selected surface governance kernels, which provides more interpretable solution trajectory behavior but hampers the agent performance.
Hence, similar to GPR, the convergence performance for a MARL task depends on the initial population of the selected governance kernels for agents.

\subsection{Performance Analysis}

\begin{figure*}
\centering
\includegraphics[width=.75\textwidth]{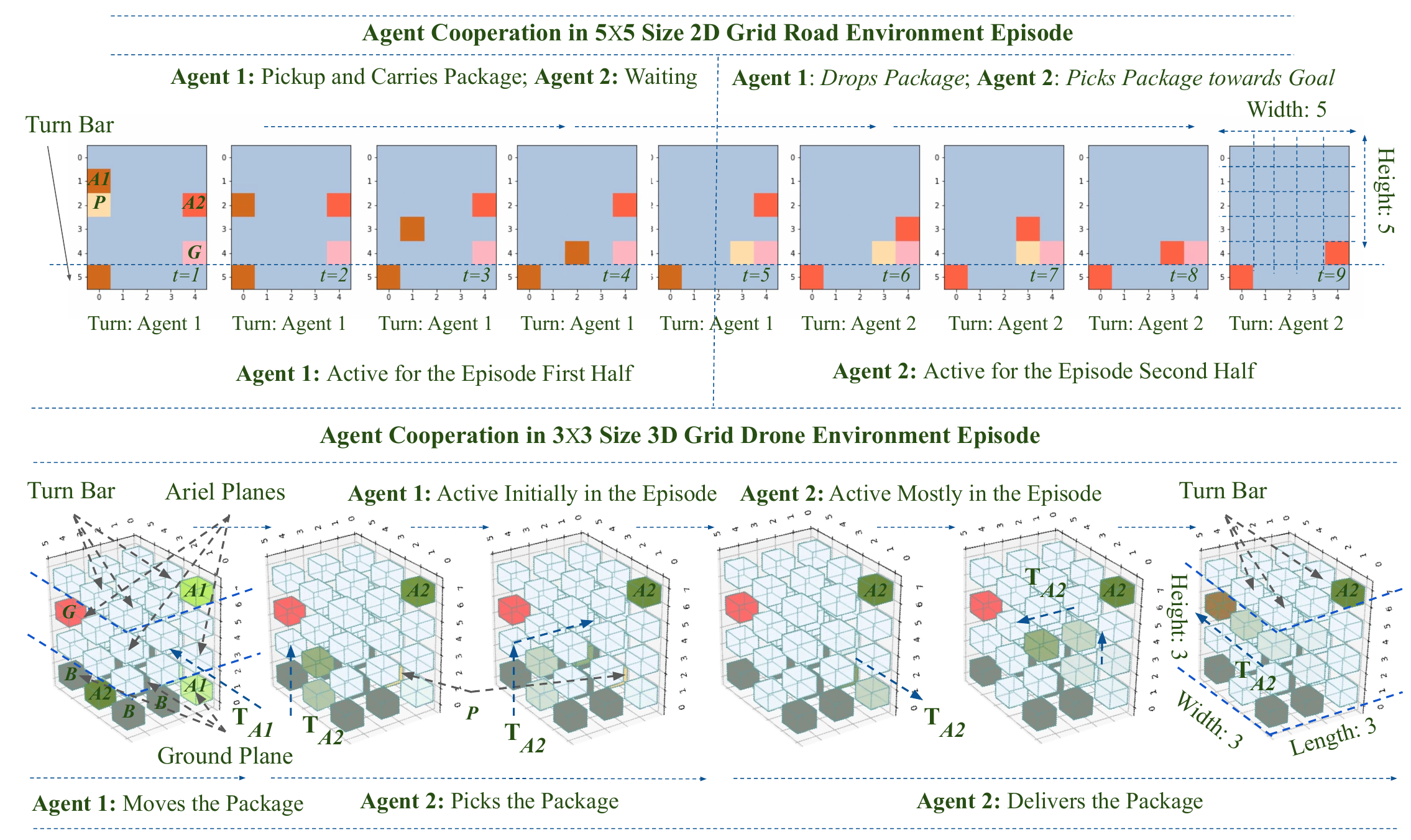}
\caption{Simulation examples for the governed cooperation behavior solution trajectories in 2D-grid road and 3D-grid drone environments for the package delivery task.}
\label{fig:governed-drone-and-grid-cooperation-trajectories}
\end{figure*}

To benchmark our approach's performance, we compare its learning behavior against the MORS approach for 120K timesteps on $5 \times 5$ 2D-grid road environments with fixed and random initial position configurations with a single goal reward ($R = 2.5$).
\footnote{The experiment implementation is available at the repository:
\href{https://github.com/arana-initiatives/gov-rek-marls}{github.com/arana-initiatives/gov-rek-marls}}
In earlier robustness and scalability experiments, the second agent proceeds with a fixed delay after the first agent starts moving.
Nevertheless, for a more realistic benchmarking comparison in this experiment, both agents move together at the episode initialization.
The reward model for the MORS approach assigns manually engineered subtask rewards like package pickups, package exchange, and proceeding closer to the goal as highlighted in Figure \ref{fig:gov-func-desc}.
From Figure \ref{fig:gov-rek-vs-mors-obj}, the governed MARLS converges relatively faster than the MORS approach, especially for randomized initial configurations.
Further, the MORS approach accumulates more reward after convergence with larger episode lengths.
In contrast, the governed MARL decreases the episode length faster during learning and more after attaining convergence.
This behavior demonstrates that the MORS approach is more prone to positive reward cycles leading to larger average episode lengths.
Hence, GOV-REK complies with the PBRS framework and multi-faceted objective to select governance kernels, minimizing average episode length and making it more fault tolerant.


\subsection{Non-spatial Problem Analysis}

Figure \ref{fig:non-spatial-intuition} demonstrates the periodic geometric trend in the joint action reward payoff matrix's flattened topology for the N-player social dilemma problem.
To extend the applicability of the GOV-REK framework, we apply all positive normalized and zero-mean normalized governance kernels on the above-described two different social dilemma problem variants.
In Figure \ref{fig:social-dilemma-result-summary}, we report the average reward accumulated in homogeneous ($r^{avg}_i$=1, $r^{max}_i$=1) and heterogeneous ($r^{avg}_i$=1.5, $r^{max}_i$=2,  $r^{min}_i$=1) scenarios with baseline and sparse payoffs.
Our experiments are carried out on a 16-agent and 16-episode length setting, where we observe that governed agents accumulate more rewards on average, especially with the zero-mean governance kernel.
\footnote{The experiment implementation is available at the repository:
\href{https://github.com/arana-initiatives/boosting-social-dilemma-collusion}{github.com/arana-initiatives/boosting-social-dilemma-collusion}}
In the homogeneous setting, the additional average rewards are accounted for by the extra reward values added by governance kernels.
However, for the heterogeneous setting, the average reward accumulation is relatively more considerable, highlighting the governed kernels' better performance in relatively more challenging settings.
Further, in sparse settings, the governed agents accumulate more rewards on average, demonstrating our approach's efficacy in sparse non-spatial \textit($S=\phi$) environments as well.

\subsection{Discussion}

Figure \ref{fig:3x3-qualitative-trajectory-plot} qualitatively demonstrates that even at small scales, the A2C algorithm with governance does not learn to deliver the package but only to exchange it.
Also, we observe that even for a small $3 \times 3$ 2D-grid road configurations, the agents are susceptible to fall for positive reward loops, which the PBRS constraint successfully handles.
Also, we highlighted the efficacy of using reward decays for the already visited states as an effective way to increase agent curiosity regions to sample diverse solution trajectories.
For the package delivery problem, the rewards are accumulated by the whole system.
Figure \ref{fig:governed-drone-and-grid-cooperation-trajectories} highlights the extent of contribution by agents in the 2D-grid road and 3D-grid drone environments.
We observe asymmetry in delivery contribution trajectories with the soft regional constraints added by our governance layer rewards.

In simpler 2D-grid road environments, we observe that the first agent contributes more in moving the package closer toward the goal, whereas in complex 3D-grid drone environments, it is vice-versa.
Figure \ref{fig:governed-drone-and-grid-cooperation-trajectories} also highlights that the first agent is more active for $5 \times 5$ 2D-grid road environment, whereas the second agent is more active for $3 \times 3$ 3D-grid drone environment.
Further, for the social dilemma problem, the contribution is more symmetric, where all agents earn similar average rewards owing to the CTDE training paradigm, where each agent maintains their policies compared to the single joint policy in CTCE package delivery task training.
We observe cooperation inconsistencies and suboptimality at larger and highly randomized configurations, which leads to larger average episode lengths and lower average reward accumulation.
However, our experiment demonstrates that our proposed approach does indeed assist the baseline MARL algorithms to achieve faster convergence without any hyperparameter tuning.


\section{Conclusion and Future Work}

Our experiments demonstrate that our proposed GOV-REK framework is robust and applies to different MARL tasks.
We demonstrate that simple additional reward model functions defined by Gaussian functions and 3D-surface functions practically help achieve faster convergence.
This paper shows that additional reward models can be successfully defined based on state or joint-action similarities for agents in a problem-agnostic manner, provided the PBRS and exploration expectation constraint is satisfied.
Further, our experimentation quantifies the practical utilization of this reward model simplification constraints for incentivizing cooperation in sparse MARL problems.

The baseline PPO-based agents at a larger scale fail to hold the exploration expectation assumption $E_a$ $\bigl[$R(s, a, $s')\bigl]$ $\rightarrow$ $R'$(s, $s'$) consistently.
Therefore, experimenting with other algorithms, like RND \cite{burda2018exploration}, NGU \cite{badia2020never}, and Agent 57 \cite{badia2020agent57}, can yield better results.
In contrast to our reward-shaping-based inductive bias, approaches like NGU and ATA attempt to learn these state similarities alongside the primary learning problem \cite{xiao2022agent}.
Thus, exploring a paradigm that trades between our rigid and simplistic reward exploration method against a wholly fluid and complex state similarity learning method is part of our future research effort.

\balance
\bibliography{references}
\end{document}